\documentclass[aps,prl,twocolumn,superscriptaddress]{revtex4} 
\usepackage{graphicx} 
\begin{document}

\title{UPPER BOUNDS ON SUSY CONTRIBUTIONS TO $b \to s$ TRANSITIONS\\
 FROM $B_s - \bar B_s$ MIXING}

\author{M.~Ciuchini}
\affiliation{Dipartimento di Fisica, Universit\`a di Roma Tre 
and INFN, Sezione di Roma III, Via della Vasca Navale 84, I-00146   
Roma, Italy}
\author{L.~Silvestrini}
\affiliation{Dipartimento di Fisica, Universit\`a di Roma ``La
  Sapienza''  and INFN, 
  Sezione di Roma, P.le A. Moro 2, I-00185 Rome, Italy}

\begin{abstract}
  We study the constraints on supersymmetric contributions to $b \to
  s$ transitions from the recent allowed range and measurement of
  $B_s-\bar B_s$ mixing obtained by the D0 and CDF collaborations at the
  Tevatron.  We compute the upper bounds on the relevant off-diagonal
  squark mass terms and compare them with the bounds coming from
  $\Delta F=1$ decays. We find that the constraints on
  chirality-flipping mass insertions are unaffected.  Conversely, the
  measurement of $B_s-\bar B_s$ mixing is effective in constraining
  chirality-conserving mass insertions, and it has striking effects in
  the case in which left- and right-handed insertions have similar
  size. Finally, we discuss the phase of the $B_s-\bar B_s$ mixing
  amplitude in the presence of SUSY contributions.
\end{abstract}

\maketitle

The study of Flavour Changing Neutral Currents (FCNC) and CP violation
at low energies is a very powerful probe of New Physics (NP). In
particular, the Unitarity Triangle (UT) analysis \cite{ckmfit,utfitNP}
and the study of rare $K$ and $B$ decays \cite{lucaLP} provide
stringent constraints on additional sources of flavour and CP
violation beyond the SM. The impact of these constraints on
supersymmetry (SUSY) is impressive \cite{pellicani}. Indeed, the
Minimal Supersymmetric Standard Model (MSSM) contains about a hundred
new sources of flavour and CP violation, mainly given by the sfermion
mass matrices \cite{Hall}. A closer look at the data from $K$ and
$B_d$ physics reveals that, while new sources of flavour violation in
$s \to d$ and $b \to d$ transitions are strongly constrained, the
possibility of large NP contributions to $b \to s$ transitions remains
open \cite{utfitNP}.

Interestingly, CP violation in nonleptonic $b \to s$ penguin decays,
such as $B \to \phi K_S$, exhibits some hint of a departure from SM
expectations \cite{bspenguinsexp}. SUSY can account for such
deviations while respecting all other available constraints from $B$
physics \cite{susybs}. In addition, in SUSY Grand Unified Theories
(GUTs) the large mixing angle observed in the neutrino sector can be
connected to a large mixing between right-handed b- and s-type
squarks \cite{bsneutrino}. Therefore, SUSY models with large
contributions to $b \to s$ transitions have received a lot of interest
recently \cite{susybsothers}. 

A crucial piece of information which has been missing until now is the
amplitude and phase of $B_s - \bar B_s$ mixing. Clearly, this would
provide an independent source of information on $b \to s$
transitions. While the $B_s - \bar B_s$ mixing phase remains unknown,
a preliminary measurement of the mixing amplitude has
been presented very recently \cite{cdf}. The aim of this paper is to
assess the impact of this measurement on SUSY sources of $b \to s$
transitions. 

To fulfill our task in a model-independent way we use the mass-insertion
approximation. Treating off-diagonal sfermion mass terms as
interactions, we perform a perturbative expansion of FCNC
amplitudes in terms of mass insertions. The lowest nonvanishing order
of this expansion gives an excellent approximation to the full
result, given the tight experimental constraints on flavour
changing mass insertions. It is most convenient to work in the
super-CKM basis, in which all gauge interactions carry the same
flavour dependence as SM ones. In this basis, we define the mass
insertions $\left(\delta^d_{ij}\right)_{AB}$ as the off-diagonal mass
terms connecting down-type squarks of flavour $i$ and $j$ and
helicity $A$ and $B$, divided by the average squark mass.

The constraints on $\left(\delta^d_{23}\right)_{AB}$ have been studied
in detail in ref.~\cite{susybs}, using as experimental input the
branching ratios and CP asymmetries of $b \to s \gamma$ and $b \to s
\ell^+ \ell^-$ decays, and the lower bound on $B_s - \bar B_s$ mixing
previously available. An update using the summer 2005 data has been
presented in ref.~\cite{lucaLP}. We perform the same analysis using,
instead of the previously available lower bound, the following result
for $B_s - \bar B_s$ mixing:
\begin{equation}
  \label{eq:dms}
  \Delta m_s = 17.33^{+0.42}_{-0.21} \pm 0.07 \mathrm{ps}^{-1}
\end{equation}
and refer the reader to \cite{susybs} for the details of the
procedure~\footnote{In a previous version of this Letter, we used the
  milder two-sided $90\%$ C.L. bound from the D0 collaboration~\cite{d0}.}.

\begin{figure*}[!htb]
\begin{center}
\includegraphics[width=0.45\textwidth]{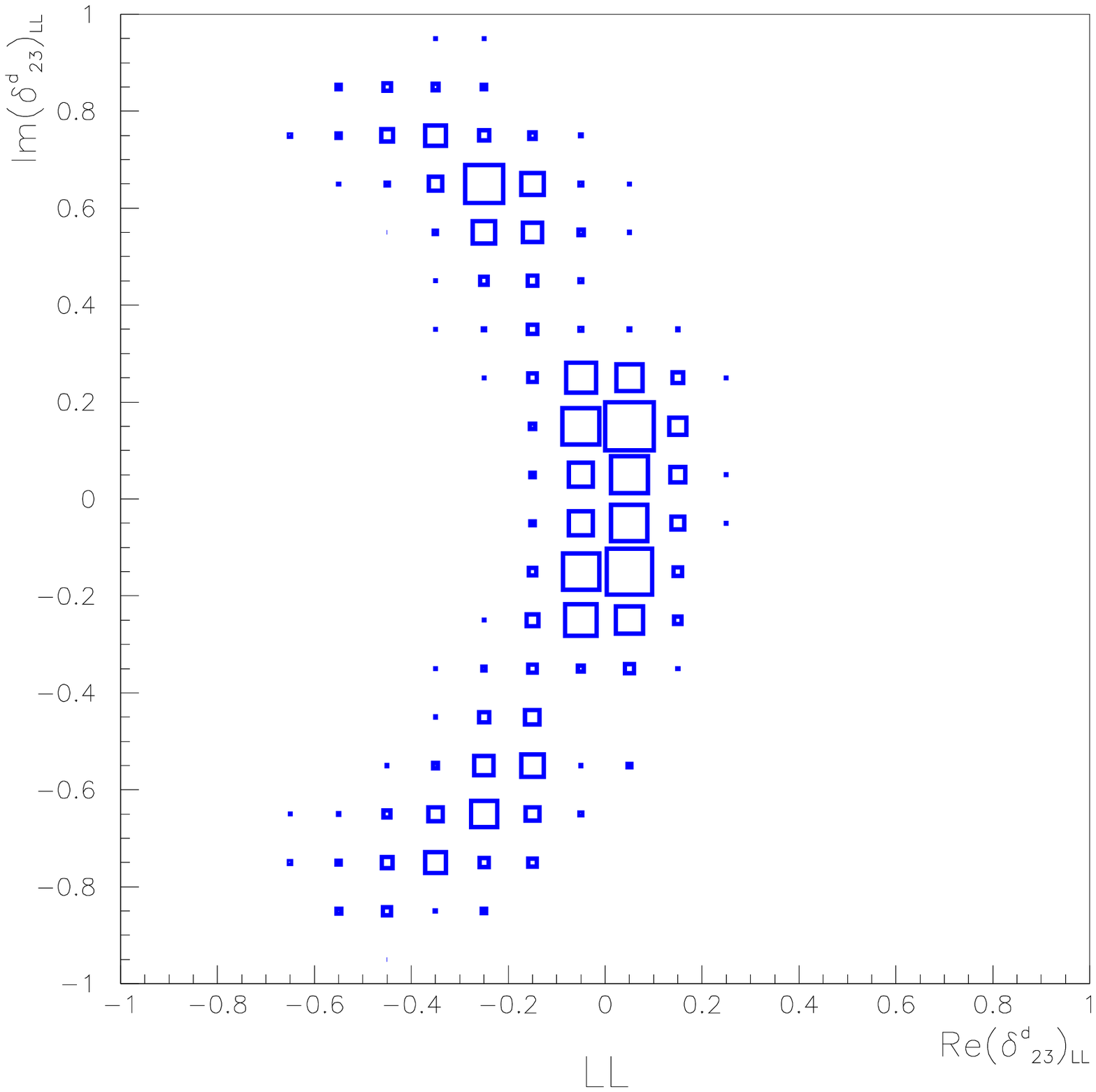} 
\includegraphics[width=0.45\textwidth]{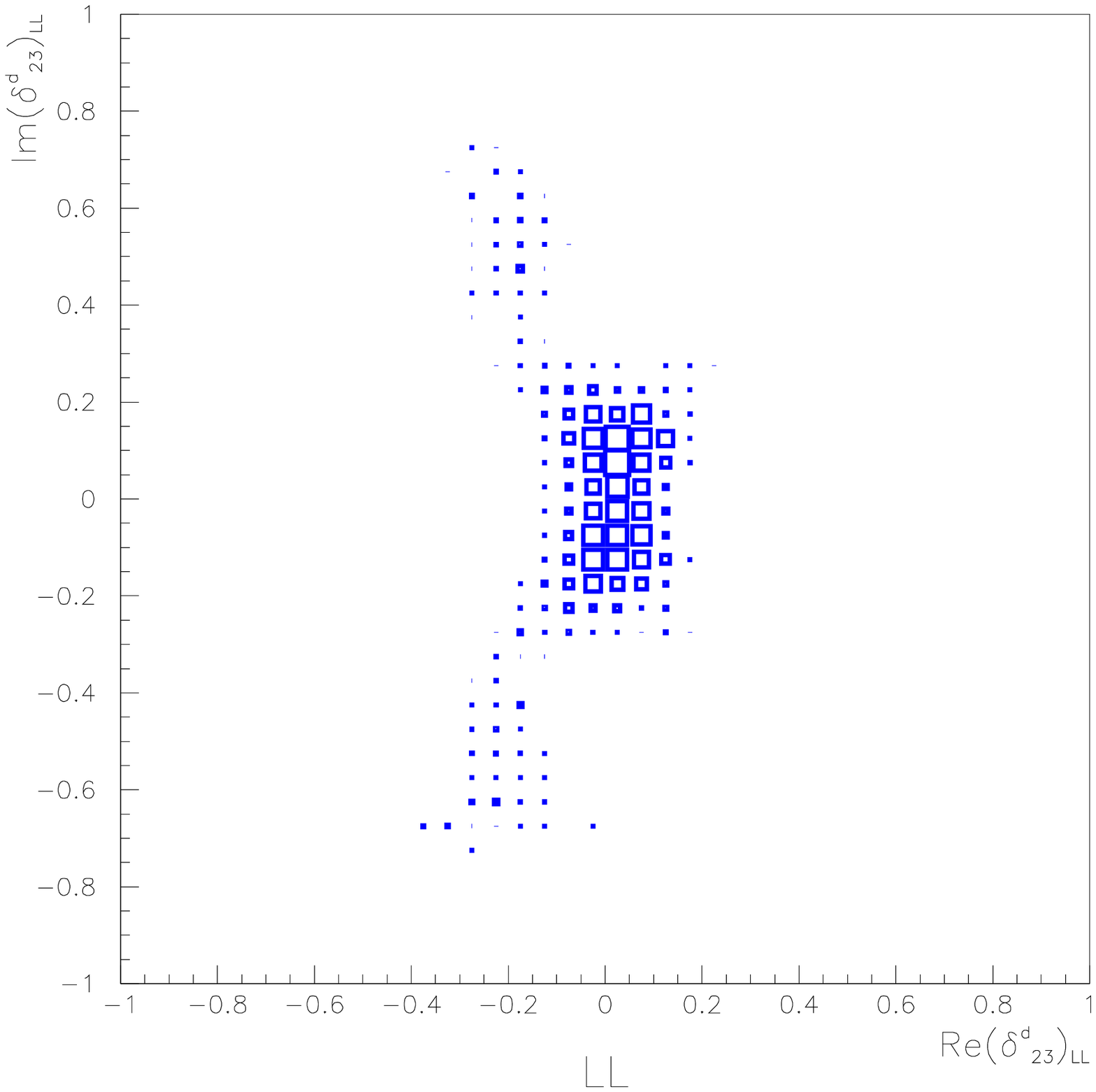} \\
\includegraphics[width=0.45\textwidth]{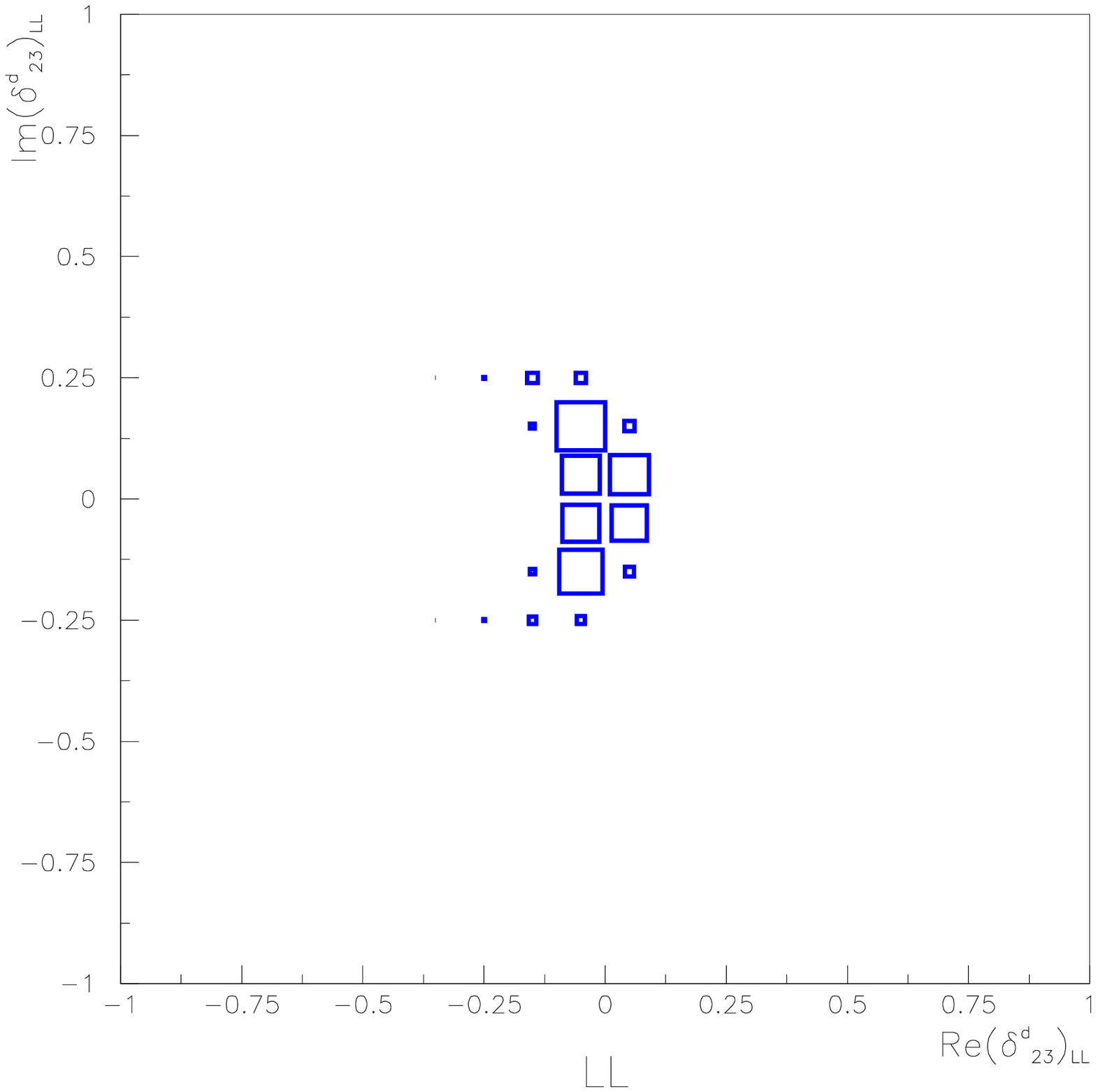} 
\includegraphics[width=0.45\textwidth]{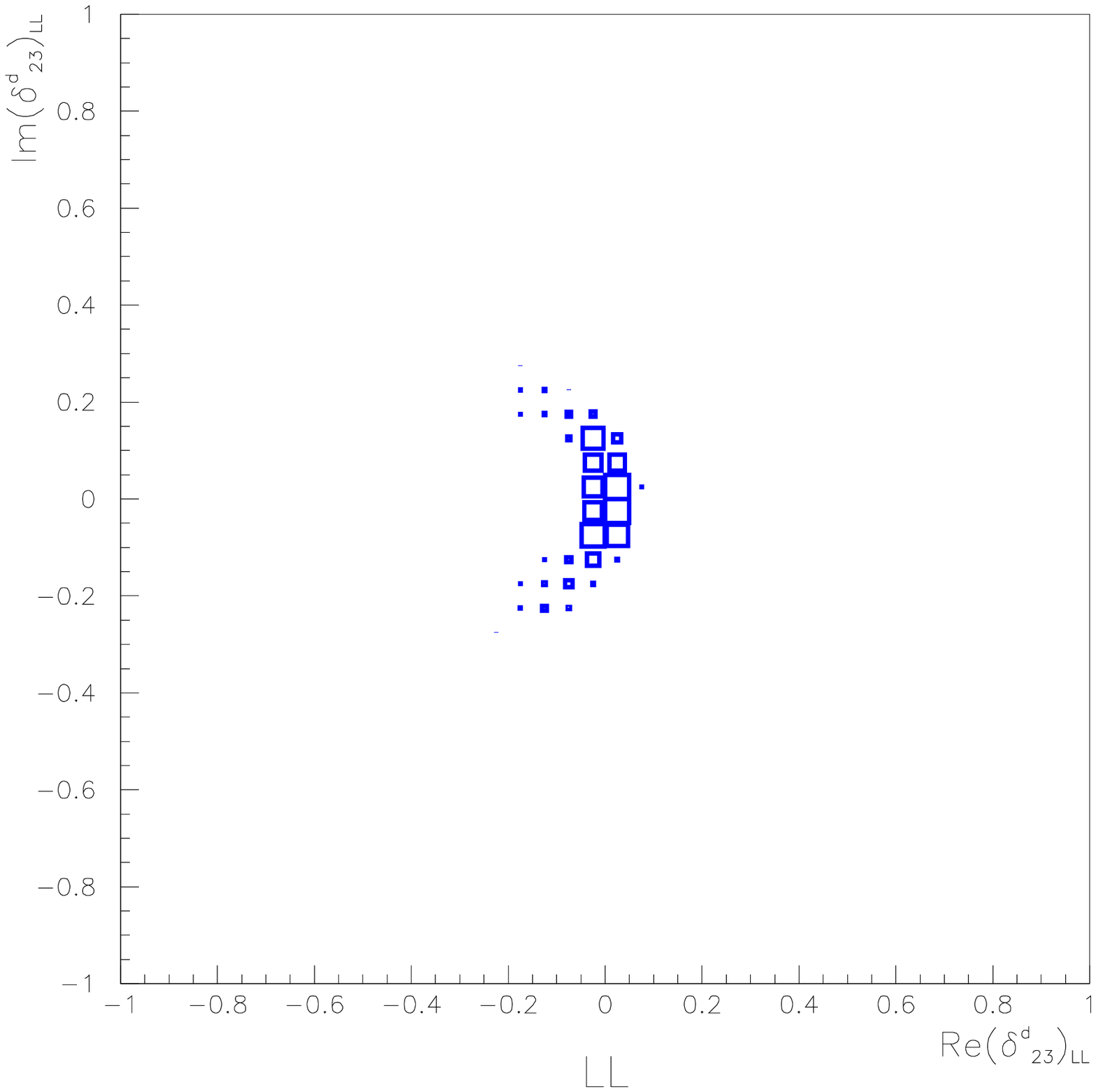}
\caption{Allowed range in the
  Re$\left(\delta^d_{23}\right)_{LL}$-Im$\left(\delta^d_{23}\right)_{LL}$
  plane. In the plots on the left (right), the lower bound
  (measurement) on $\Delta m_s$ is used. Plots in the upper (lower)
  row correspond to $\tan\beta=3$ ($\tan\beta=10$). See the text for
  details.}
\label{fig:LL}
\end{center}
\end{figure*}

\begin{figure*}[!htb]
\begin{center}
\includegraphics[width=0.45\textwidth]{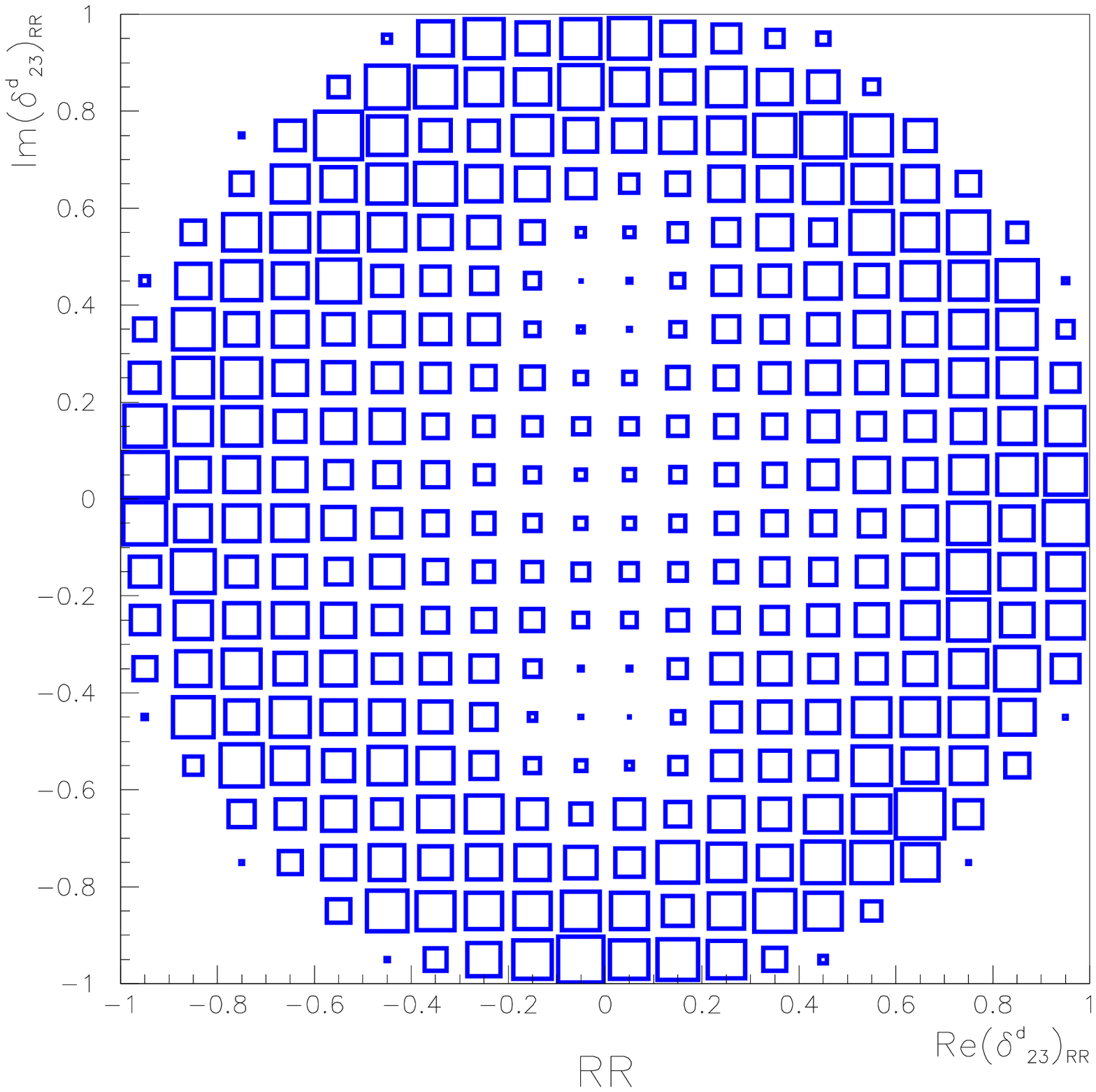} 
\includegraphics[width=0.45\textwidth]{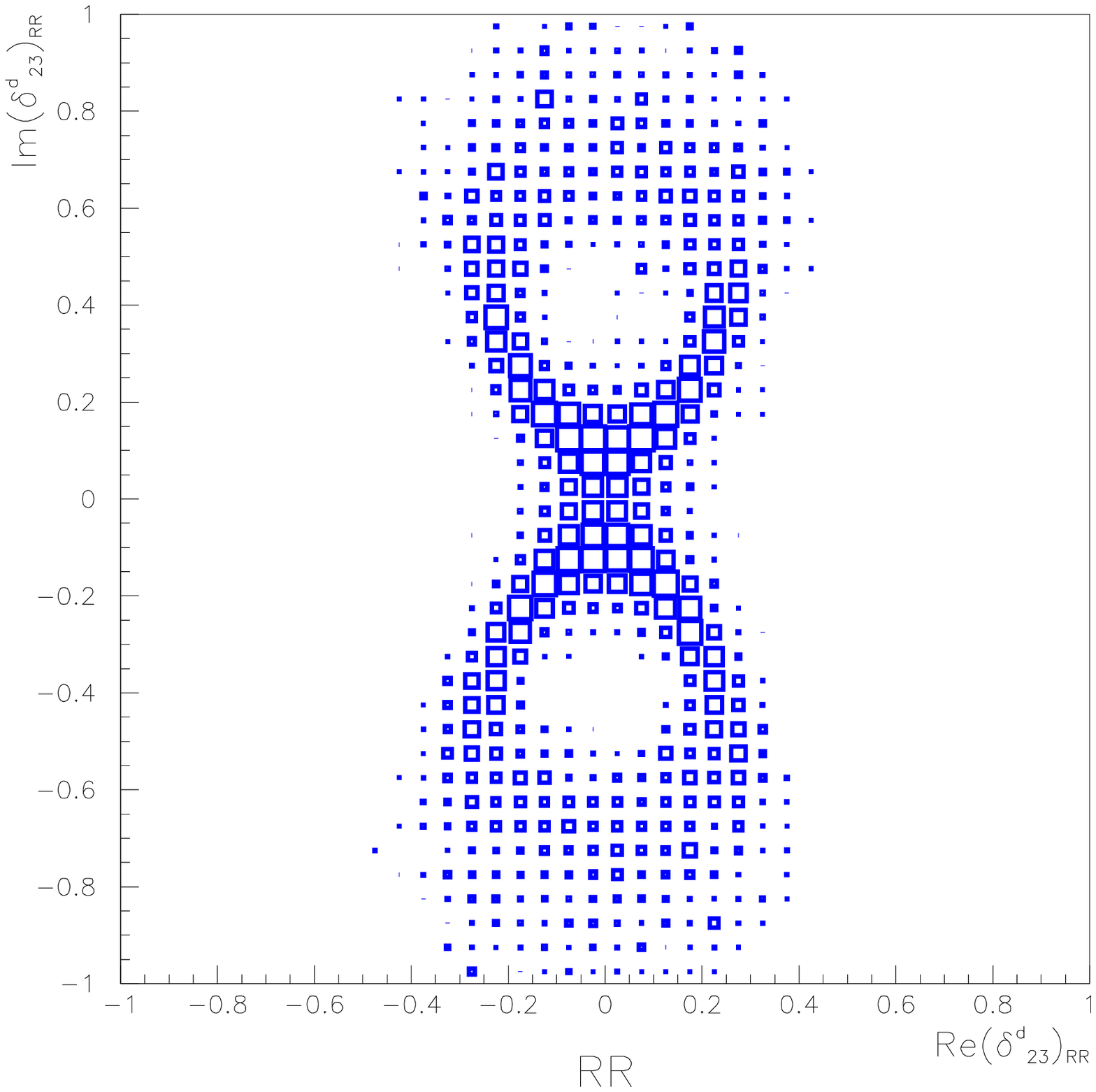} 
\caption{Allowed range in the
  Re$\left(\delta^d_{23}\right)_{RR}$-Im$\left(\delta^d_{23}\right)_{RR}$
  plane. In the plots on the left (right), the lower bound
  (measurement) on $\Delta m_s$ is used. See the text for details.}
\label{fig:RR}
\end{center}
\end{figure*}

\begin{figure*}[!htb]
\begin{center}
\includegraphics[width=0.45\textwidth]{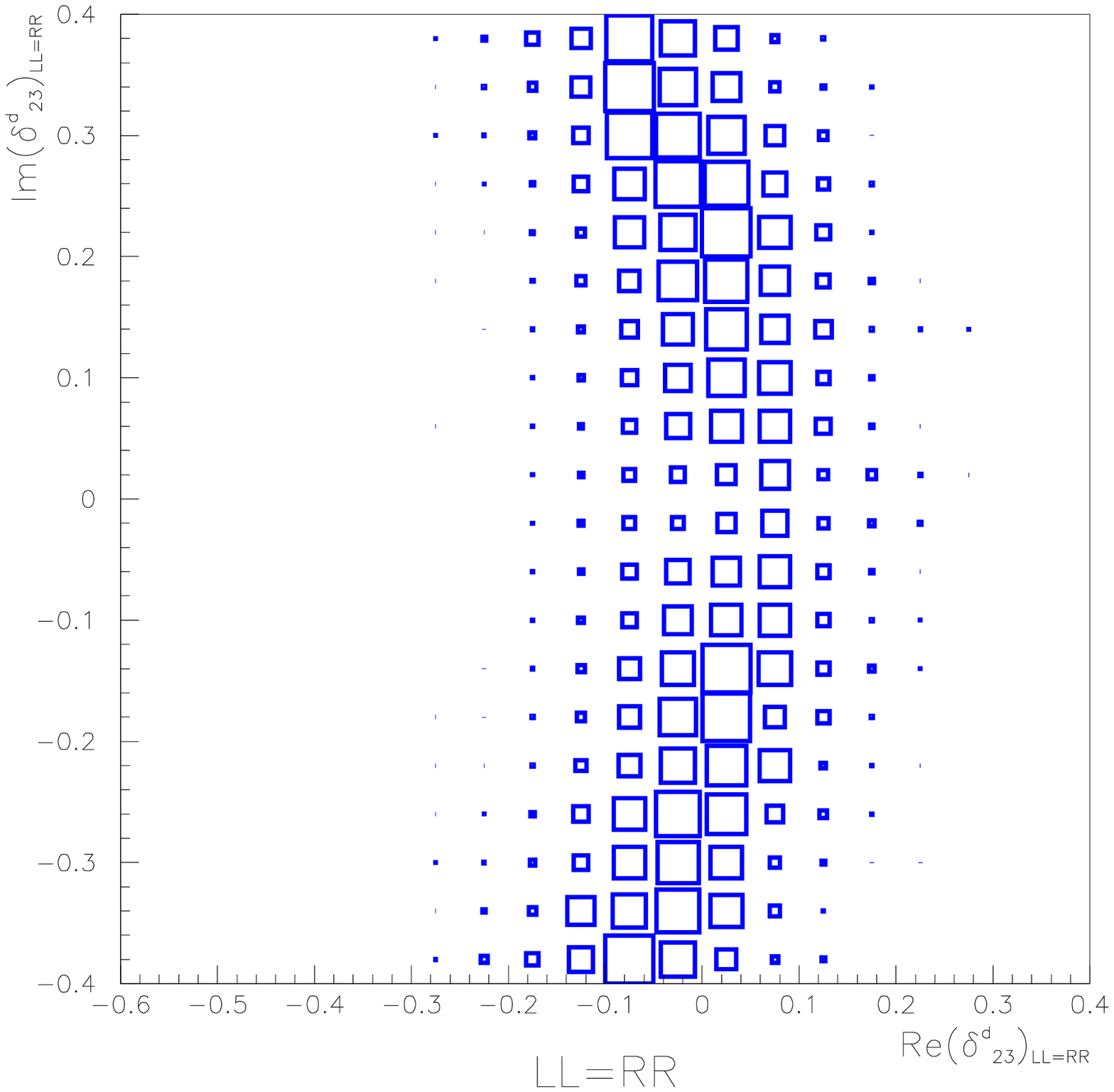} 
\includegraphics[width=0.45\textwidth]{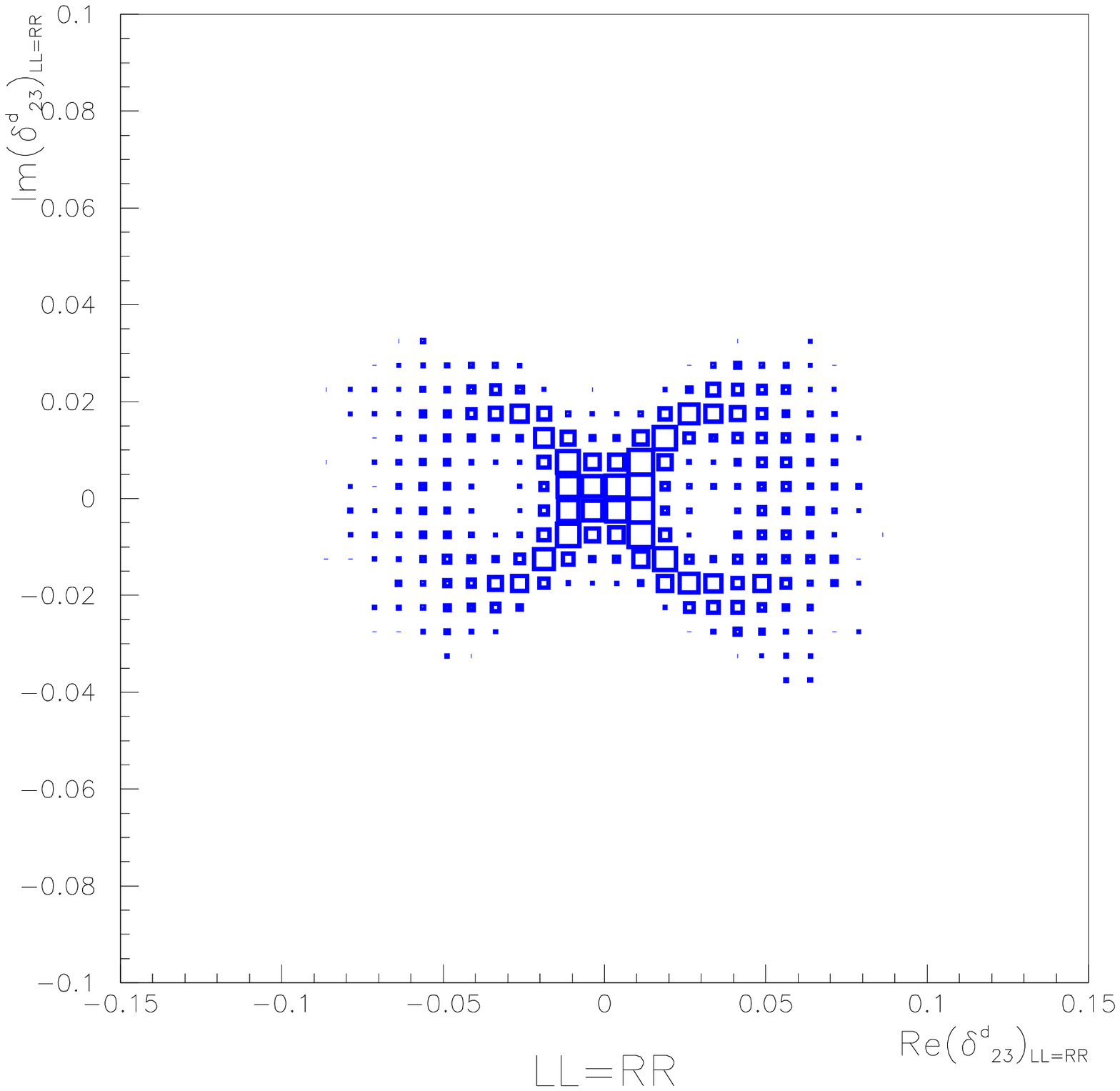} 
\caption{Allowed range in the
  Re$\left(\delta^d_{23}\right)_{LL=RR}$-Im$\left(\delta^d_{23}\right)_{LL=RR}$
  plane. In the plots on the left (right), the lower bound
  (measurement) on $\Delta m_s$ is used. See the text for details.}
\label{fig:LLRR}
\end{center}
\end{figure*}

For definiteness, we choose an average squark mass of $350$ GeV, a
gluino mass of $350$ GeV, $\mu=-350$ GeV and $\tan
\beta=3$~\footnote{The dependence on $\mu$ and on $\tan \beta$ is
  induced by the presence of a chirality flipping, flavour conserving
  mass insertion proportional to $\mu\tan \beta$.}. In
Fig.~\ref{fig:LL}, we present the allowed range in the
Re$\left(\delta^d_{23}\right)_{LL}$-Im$\left(\delta^d_{23}\right)_{LL}$
plane, using the previous lower bound (upper left) or the present
measurement (upper right) of $\Delta m_s$. We see that the effect of
the measurement is to suppress the regions
$\vert$Im$\left(\delta^d_{23}\right)_{LL}\vert\gtrsim 0.2$. For higher
values of $\tan \beta$, the effect of $\Delta m_s$ becomes less
important, since the constraints from $\Delta F=1$ processes become
comparable for $\tan \beta\sim 10$ and then dominate, as can be seen
looking at the lower part of Fig.~\ref{fig:LL}, where the same
analysis has been performed with $\tan \beta=10$.

In Fig.~\ref{fig:RR}, we present the allowed range in the
Re$\left(\delta^d_{23}\right)_{RR}$-Im$\left(\delta^d_{23}\right)_{RR}$
plane, using the previous lower bound (left) or the present value
(right) of $\Delta m_s$ (the plots correspond to $\tan \beta=3$, but
the dependence on $\tan \beta$ is negligible here and in the
following). The effect of the constraint from $\Delta m_s$ can clearly
be seen, leading to an upper bound on
$\vert$Re$\left(\delta^d_{23}\right)_{RR}\vert$ around $0.4$. The
effect of the bound from $\Delta m_s$ is much more striking in the
case of
$\left(\delta^d_{23}\right)_{LL}=\left(\delta^d_{23}\right)_{RR}$, as
can be seen in Fig.~\ref{fig:LLRR}. In this case, the upper bound
obtained on $\vert\left(\delta^d_{23}\right)_{LL=RR}\vert$ is around
$10^{-1}$. For the other mass insertions
$\left(\delta^d_{23}\right)_{LR,RL}$, the constraint from $\Delta m_s$
is irrelevant since the main effect comes from $\Delta F=1$ processes.
This leaves open the possibility of sizable deviation from the SM
prediction in the CP asymmetries in $b \to s$ penguin decays
\cite{lucaLP}.

\begin{figure*}[!htb]
\begin{center}
\includegraphics[width=0.45\textwidth]{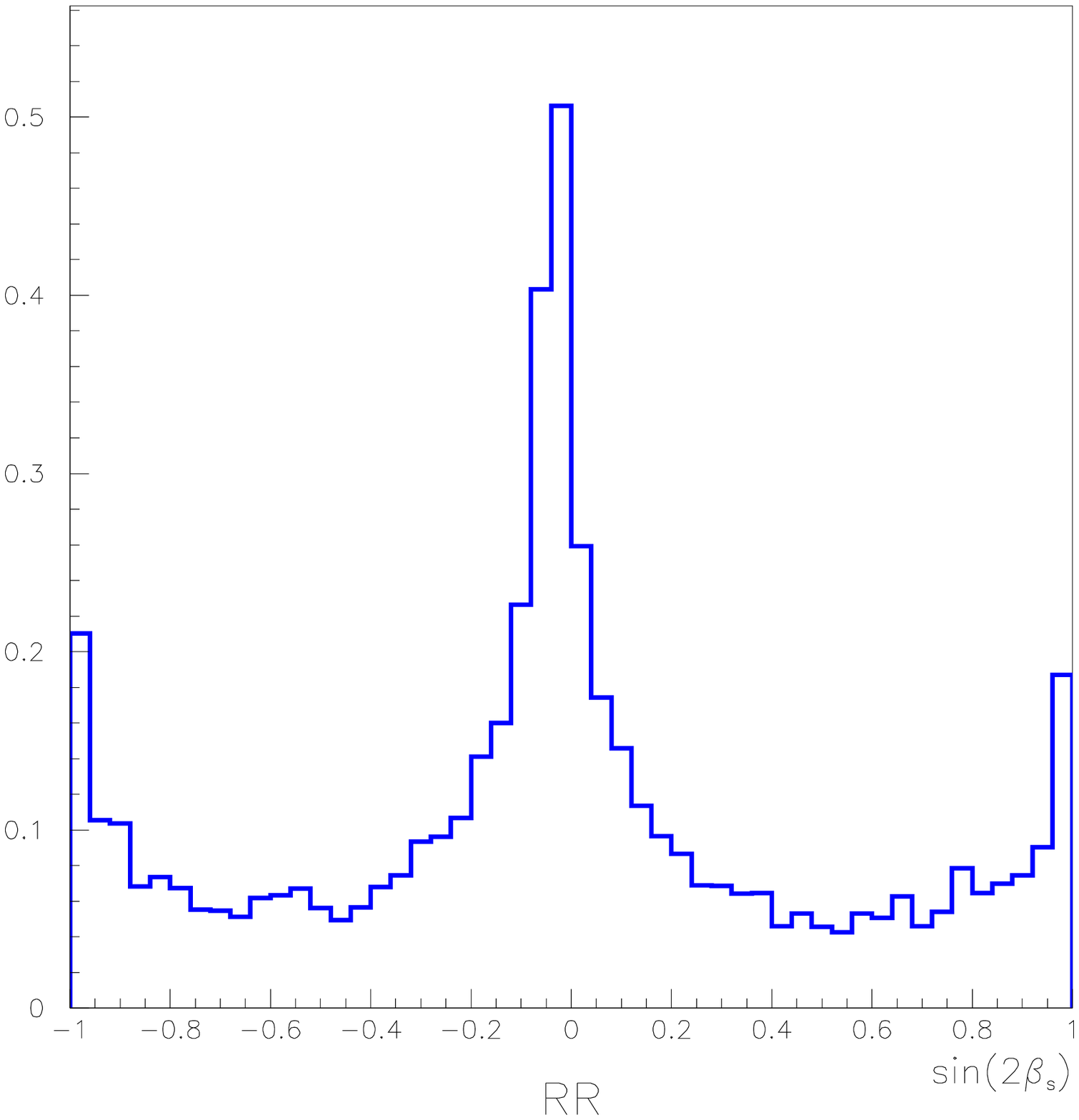} 
\includegraphics[width=0.45\textwidth]{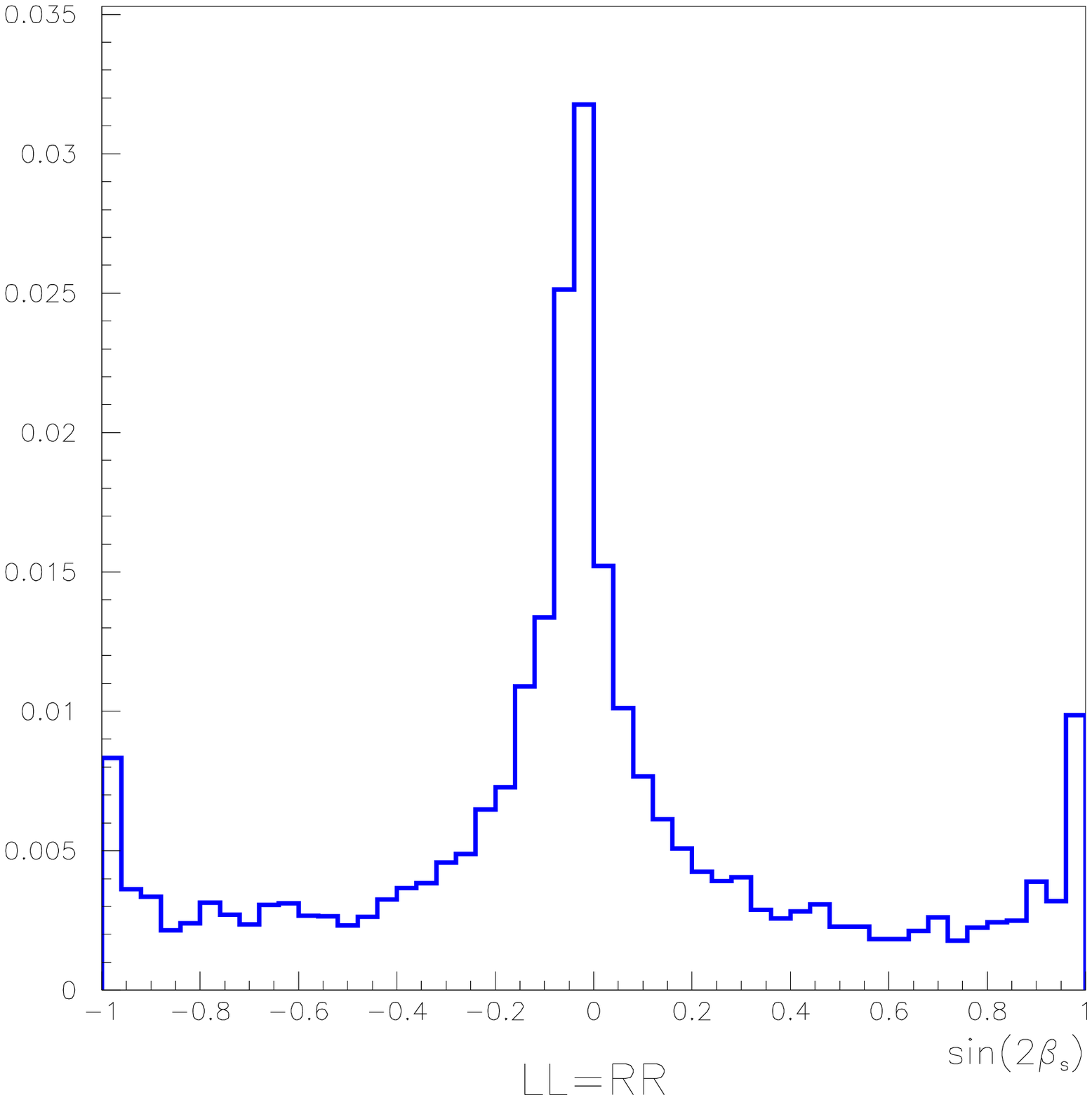} 
\caption{P.d.f. for $\sin 2 \beta_s$ in the presence of
  $\left(\delta^d_{23}\right)_{RR}$ (left) or
  $\left(\delta^d_{23}\right)_{LL=RR}$ (right). See the text for details.}
\label{fig:s2bs}
\end{center}
\end{figure*}

Finally, in Fig.~\ref{fig:s2bs} we present the p.d.f. for $\sin 2
\beta_s$, where $2 \beta_s$ is the phase of the $B_s-\bar B_s$ mixing
amplitude, relevant for measurements of CP violation in $B_s$ physics.
Comparing these results with the SM value $\sin 2 \beta_s^\mathrm{SM}=-(3.65 \pm
0.30)\times 10^{-2}$ \cite{utfitSM}, it is evident that there is still
plenty of room to observe non-standard CP violation in the $B_s$
system.

This work has been supported in part by the EU network ``The quest for
unification'' under the contract MRTN-CT-2004-503369.

\end{document}